\magnification=\magstep1
\overfullrule=0pt
\hsize=15.4truecm
{\nopagenumbers \line{\hfil UQAM-PHE-95/14}
\vskip.1cm
\centerline{\bf GLUINO AND NEUTRALINO CONTRIBUTION TO THE DIRECT} 
\centerline{\bf CP VIOLATION PARAMETER $\epsilon'$}
\vskip2cm
\centerline{{HEINZ K\"ONIG}
\footnote*{email:konig@osiris.phy.uqam.ca}}
\centerline{D\'epartement de Physique}
\centerline{Universit\'e du Qu\'ebec \`a Montr\'eal}
\centerline{C.P. 8888, Succ. Centre Ville, Montr\'eal}
\centerline{Qu\'ebec, Canada H3C 3P8}
\vskip3cm
\centerline{\bf ABSTRACT}\vskip.2cm
\noindent
I present a detailed and complete calculation of the gluino
and neutralino contribution to the direct CP violating parameter 
$\epsilon'$\ within the MSSM.
I include the complete mixing matrices of the neutralinos 
and of the scalar partners of the left and right
handed down quarks. 
I find that the neutralino contribution is generally small but 
can be larger than 
the gluino contribution for small values $m_S\le 400$\ GeV
of the supersymmetric breaking scale. 
\vskip2cm
\centerline{ December 1995} 
\vfill\break}
\pageno=1
\noindent
{\bf I. INTRODUCTION}\hfill\break\vskip.2cm\noindent
Although the standard model (SM) predicts observable
levels of direct CP violation in the K system for a 
large top quark mass, there is still a disagreement
of the experimental results. Whereas the one of the
NA31 collaboration at CERN [1] is given by $Re\bigl (\epsilon'/
\epsilon\bigr )=23\pm 6.5\times 10^{-4}$, the collaboration
E731 at Fermilab [2] reports a value of  $Re\bigl (\epsilon'/
\epsilon\bigr )=7.4\pm 6.0\times 10^{-4}$. For a top quark mass
of $m_t=180\pm 12$\ GeV [3] the SM predicts $\epsilon'/\epsilon$\
to be in the range of $(0-3)\times 10^{-3}$\ [4--7], where
the uncertainty lies among cancellations of strong and
electroweak penguin diagrams, and thus is still in agreement with
both experiments.\hfill\break\indent
It is therefore of great interest to look at the contributions
to this direct CP violating parameter in any kind of models 
beyond the SM of which the most promising nowadays is its
minimal supersymmetric extensions (MSSM) [8]. Studies of the
contribution to $\epsilon'/\epsilon$\ within this model were
done a while ago [9--12 and references therein]. More recent
work can be found in [13, 14 and references therein].
\hfill\break\indent
Since in the $K^0$\ system
$(K_{21}^\ast K_{22})^2m^2_c\ge (K^\ast_{31}K_{32})^2m^2_t$\
the charm quark contribution is larger than the top quark
contribution, whereas it is vice versa in the $B^0_d$\
system.
In recent papers [15, 16] we have
shown that the largest contribution to
the mass difference in the $B^0_d$\ system within
the MSSM were given by the charged Higgses for small
$\tan\beta=v_2/v_1$\ (with $v_{1,2}$\ the vacuum expectation
values, vev's, of the neutral Higgs bosons) and the charginos.
The gluino contributes non negligibly when its mass is
small, of order $100$\ GeV, and the neutralino contribution
was shown to be non neglectable for large $\tan\beta\approx 50$\
and small values of the SUSY breaking mass parameter $m_S\le 300$\
GeV.\hfill\break\indent 
In the literature the SUSY contribution to the 
 mass difference of the $K_0$\ system 
and therefore to the parameter
$\epsilon$\ (the imaginary part of the matrix elements 
leading to the mass difference) was supposed
to be neglectable. As a consequence emphasis was put in
calculating the penguin diagrams leading to $\epsilon'$.
However in [13, 14] the authors found non neglible interference
between the box and penguin diagrams for certain ranges of
$x=m^2_{\tilde g}/m_S^2$\ in the proximity of $1$. 
\hfill\break\indent
In this paper I do a reanalysis of the penguin diagrams
as shown in Fig.1 with gluinos and scalar down quarks
within the loop, which contribute to the direct CP 
violation parameter $\epsilon'$, reproducing the results
of [9]\footnote{$^1$}{I therefore do not agree with the
statement of [17] (analysing the contribution to the decay
$b\rightarrow sg$\ within the MSSM), that the authors
of [9] have neglected a crucial term}. In the calculations
I also include the neutralino contribution. I do not neglect
any mixing of the scalar partners of the left and right 
handed down quarks nor any mixing in the neutralino
sector. The goal of this paper is to compare the gluino
contribution with the neutralinos
one and to show that the latter in general can not be
neglected. Here I do not give a complete
analysis of the contribution of all particles within the
MSSM (as there are the charged Higgs boson and charginos)
or all sort of diagrams (including the box diagrams), which
will be presented elsewhere [18]. 
\hfill\break\indent
In the next section I present the calculation and discuss the
results in the third section. I end with the conclusions.
Since in the literature [17] the correctness of the results
in [9] was doubted I present the detailed calculation in the
appendices.
\hfill\break\vskip.2cm\noindent
{\bf II. GLUINO AND
NEUTRALINO CONTRIBUTION TO THE DIRECT CP VIOLATING
PARAMETER $\epsilon'$}\vskip.2cm  
In the SM there are two CP violation parameter: the indirect
CP violation parameter $\epsilon$\ and the direct CP violation
parameter $\epsilon'$. Whereas the indirect one follows 
from the mass eigenstates of $K^0$\
and $\overline K^0$\ and is given by the imaginary
parts of the diagrams leading to $\Delta m_{K^0}$, $\epsilon'$\ 
describes the direct decay of the Kaons into two pions.
The values are given by:
$$\eqalignno{\epsilon=&{{e^{i\pi/4}}\over{\sqrt{2}{\Delta m_{K^0}}}}
Im M_{12}&(1)\cr
\epsilon'\approx&-{{\omega}\over{\sqrt{2}}}\xi(1-\Omega)
e^{i\pi/4}&(2)\cr}$$
$\Delta m_{K^0}=2 Re M_{12}$, $\omega=Re A_2/Re A_0
\approx 1/22$, $\xi=Im A_0/
Re A_0$\ and $\Omega=Im A_2/\omega Im A_0$, 
where $A_0$\ and $A_2$\ are the amplitudes of the decays
into two pions with $\Delta I=1/2$\ and $\Delta I=3/2$\ respectevly.
The uncertainty in the SM lies among certain cancellations
between the electroweak and strong diagrams leading to
the $\Omega$\ term, which
approaches $1$\ for $m_t\approx 220$\ GeV and thus 
$\epsilon'/\epsilon$\ obtaines even negative values
for higher top quark masses [6,7]. In [5]
however the authors using chiral pertubation theory
obtain positive values for all top quark masses.
\hfill\break\indent
To obtain the gluino and neutralinos contribution
to the $\epsilon'$\ parameter we have to calculate
the diagrams as shown in Fig.1. In the calculation
I consider the full mass matrices of the scalar
down quarks including $1$\ loop corrections and
the mixing terms of scalar partners of the left and
right down quarks; that is I present the results
in the mass eigenstates and not in the current eigenstates
of the scalar down quarks. I also consider
the full $4\times 4$\ matrix of the neutralinos and calculate
its mass eigenstates and mixing angles numerically.
For a detailed discreption of the mass eigenstates,
mixing angles of the scalar down quarks and neutralinos
I refer the interested reader to [15, 16] and will not
represent them here. Their couplings to the quarks can
be found in Fig.24 in [19].
\hfill\break\indent
For the couplings of the gluinos to the gluons
and to the quarks and scalar quarks 
it was shown a while ago, that there occur
flavour changing strong interactions between the gluino,
the left handed quarks, and their supersymmetric
scalar partners, whereas the couplings of the gluino
to the right handed quarks and their partners remains
flavour diagonal [20, 21]. However in general this might not
be the case and therefore the authors in [9] 
took both couplings to be flavour non diagonal, which 
leads to a term, which contributes to $\epsilon'$\
proportional to the gluino mass $m_{\tilde g}$.
As was pointed out in appendix B of [19] also the couplings
of the neutralinos to the left- and right handed quarks
and their superpartners are in general flavour non diagonal. To
compare the neutralino contribution to $\epsilon'$\
with the gluino one 
I therefore include both couplings in the calculation. 
\hfill\break\indent
The Lagrangians, which describe the couplings of the gluino
to the gluon, of the gluon to the scalar down quarks and 
the flavour non diagonal one of the gluino 
to the left and right handed down quarks and their superpartners,
needed for the calculation of the penguin diagrams, are
given in the mass eigenstates of the scalar down quarks by:
$$\eqalignno{{\cal L}_{g\tilde g\tilde g}=&{i\over 2}g_sf_{abc}
\overline{\tilde g_a}\gamma_\mu\tilde g_b G^\mu_c&(3)\cr
{\cal L}_{\tilde d\tilde d\tilde g}=&-ig_sT^aG^{a\mu}\sum
\limits_{m=1,2}\tilde d_m^\ast
{\buildrel \leftrightarrow \over\partial}_\mu\tilde d_m&(4)\cr
{\cal L}_{FC}=&-\sqrt{2}g_sT^a\sum\limits_{m=1,2}\bigl\lbrack
K^{\tilde g}_L\overline{
\tilde g_a}P_Ld\tilde K^d_{m1}\tilde d^\ast_m
-K^{\tilde g}_R\overline{\tilde g_a}P_R d
\tilde K^d_{m2}\tilde d^\ast_m
\bigr\rbrack
+h.c.
&(5)\cr}$$
Eq.(3) has to be multiplied by 2 to
obtain the Feynman rules. $f_{abc}$\ are the $SU(3)_C$\
structure constants, $T^a$\ its generators and $g_s$\
the strong coupling constant.
\hfill\break $P_L=(1-\gamma_5)/2$\ and $P_L=(1+\gamma_5)/2$\
are the left and right handed helicity operators respectively. 
$K^{\tilde g}_{L,R}$\ is the
supersymmetric version of the Kobayashi--Maskawa (KM) matrix. 
 $\tilde K^d_{nm}$\ 
is the matrix descriping the relation of the mass eigenstates of
$\tilde d_{1,2}$\ to the current eigenstates 
$\tilde d_{L,R}$\ ($d$\ stands for down, strange and
bottom quarks, generation indices have been omitted)
and is given by:
$$\left(\matrix{\tilde d_1\cr \tilde d_2\cr}\right)=\left(\matrix{
\cos\Theta_d&\sin\Theta_d\cr-\sin\Theta_d&\cos\Theta_d\cr}
\right)\left(\matrix{\tilde d_L\cr \tilde d_R\cr}\right)
\eqno(6)$$ 
Following the philosophy of [9] I obtain after a lenghty
but straightforward calculation of the penguin diagrams
of Fig.1 the following results for the $\xi$\ parameter:
$$\eqalignno{\xi=&-f Im M/Re M&(7)\cr
M=&M_{SM}+M_{\tilde g}+M_{\tilde N}\cr
M_{SM}=&\sum\limits_{a=1-3}K^\ast_{a1}K_{a2}
{{\alpha}\over{2\sin^2\Theta_Wm_W^2}}\left(A_a^{SM}+
\eta_{QCD}B_a^{SM}\right)\cr
M_{\tilde g}=&-\sum\limits_{a=1-3}
{{\alpha_s}\over{m_{\tilde g}^2}}
\left(A_a^{\tilde g}+\eta_{QCD}B_a^{\tilde g}\right)\cr
M_{\tilde N}=&-\sum\limits_{a=1-3}\sum\limits_{i=1-4}
{{\alpha}\over{2\sin^2\Theta_W\cos^2\Theta_W
m^2_{\tilde N_i}}}\left(A_a^{\tilde N_i}-\eta_{QCD}B_a^{\tilde N_i}
\right)\cr
A_a^{SM}=&{2\over 3}(1+y_a)(1-y_a-{11\over 4}y_a^2-{3\over 4}
y_a^3)\log({{m_{u_a}^2}\over{m_W^2}})-{3\over 2}y_a(1+{25\over 18}
y_a+{1\over 3}y_a^2)\cr
B_a^{SM}=&\bigl\lbrack
{3\over 2}y_a^2(1+y_a)^2\log({{m_{u_a}^2}\over{m_W^2}})
+{1\over 2}y_a(1+{9\over 2}y_a+3y_a^2)\bigr\rbrack (m_s-m_Q)\cr
y_a=&{{m_{u_a}^2}\over{m_W^2-m_{u_a}^2}}\cr
A_a^{\tilde g}=&\sum\limits_{m=1,2}\bigl\lbrack
C_{2G}A_G(x_{\tilde a_m}^{\tilde g})+
2C_{2F}A_F(x_{\tilde a_m}^{\tilde g})
\bigr\rbrack\bigl\lbrack
K^{\ast\tilde g}_{a1L}K^{\tilde g}_{a2L}\tilde K^{a2}_{m1}
+K^{\ast\tilde g}_{a1R}K^{\tilde g}_{a2R}\tilde K^{a2}_{m2}
\bigr\rbrack\cr
B_a^{\tilde g}=&\sum\limits_{m=1,2}
\Bigl\lbrace\bigl\lbrack
C_{2G}B_G(x_{\tilde a_m}^{\tilde g})
-2C_{2F}B_F(x_{\tilde a_m}^{\tilde g})
\bigr\rbrack\bigl\lbrack 
K^{\ast\tilde g}_{a1L}K^{\tilde g}_{a2L}
\tilde K^{a2}_{m1}-K^{\ast\tilde g}_{a1R}K^{\tilde g}_{a2R}
\tilde K^{a2}_{m2}\bigr\rbrack
\cr&\times (m_s-m_Q)\cr
&-\bigl\lbrack C_{2G}\tilde B_G(x_{\tilde a_m}^{\tilde g})
-2C_{2F}\tilde B_F(x_{\tilde a_m}^{\tilde g})
\bigr\rbrack m_{\tilde g}\bigl\lbrack
K^{\ast\tilde g}_{a1L}K^{\tilde g}_{a2R}
-K^{\ast\tilde g}_{a1R}K^{\tilde g}_{a2L}\bigr\rbrack
\tilde K^a_{m1}\tilde K^a_{m2}\Bigr\rbrace\cr
A_a^{\tilde N_i}=&\sum\limits_{m=1,2}
A_F(x_{\tilde a_m}^{\tilde N_i})\Bigl\lbrace
K^{\ast\tilde N_i}_{a1L}K^{\tilde N_i}_{a2L}\tilde K^{a2}_{m1}
(T_i^{dL}T_i^{sL}+T_{m_di}T_{m_si})\cr
&+K^{\ast\tilde N_i}_{a1R}K^{\tilde N_i}_{a2R}
\tilde K^{a2}_{m2}(T_i^{dR}T_i^{sR}+T_{m_di}T_{m_si})
\cr&
+\tilde K^a_{m1}\tilde K^a_{m2}\bigl\lbrace
K^{\ast\tilde N_i}_{a1L}K^{\tilde N_i}_{a2R}
(T_i^{dL}T_{m_si}+T_i^{sR}T_{m_di})
\cr&
+K^{\ast\tilde N_i}_{a1R}K^{\tilde N_i}_{a2L}
(T_i^{sL}T_{m_di}+T_i^{dR}T_{m_si})\bigr\rbrace
\Bigr\rbrace\cr
B_a^{\tilde N_i}=&\sum\limits_{m=1,2}
\Bigl\lbrace
B_F(x_{\tilde a_m}^{\tilde N_i})\Bigl\lbrack
K^{\ast\tilde N_i}_{a1L}K^{\tilde N_i}_{a2L}\tilde K^{a2}_{m1}
(T_i^{dL}T_i^{sL}-T_{m_di}T_{m_si})\cr
&-K^{\ast\tilde N_i}_{a1R}K^{\tilde N_i}_{a2R}
\tilde K^{a2}_{m2}(T_i^{dR}T_i^{sR}-T_{m_di}T_{m_si})
\cr&
+\tilde K^a_{m1}\tilde K^a_{m2}\bigl\lbrace
K^{\ast\tilde N_i}_{a1L}K^{\tilde N_i}_{a2R}
(T_i^{dL}T_{m_si}-T_i^{sR}T_{m_di})
\cr&
+K^{\ast\tilde N_i}_{a1R}K^{\tilde N_i}_{a2L}
(T_i^{sL}T_{m_di}-T_i^{dR}T_{m_si})\bigr\rbrace
\Bigr\rbrack
(m_s-m_Q)\cr
&+\tilde B_F(x_{\tilde a_m}^{\tilde N_i})\Bigl\lbrack
K^{\ast\tilde N_i}_{a1L}K^{\tilde N_i}_{a2L}
\tilde K^{a2}_{m1}(T_{m_si}T_i^{dL}-T_{m_di}T_i^{sL})\cr
&-K^{\ast\tilde N_i}_{a1R}K^{\tilde N_i}_{a2R}
\tilde K^{a2}_{m2}(T_{m_si}T_i^{dR}-T_{m_di}T_i^{sR})
\cr&
+\tilde K^a_{m1}\tilde K^a_{m2}\bigl\lbrace
K^{\ast\tilde N_i}_{a1L}K^{\tilde N_i}_{a2R}
(T_i^{dL}T_i^{sR}-T_{m_di}T_{m_si})
\cr&
+K^{\ast\tilde N_i}_{a1R}K^{\tilde N_i}_{a2L}
(T_{m_si}T_{m_di}-T_i^{sL}T_i^{dR})\bigr\rbrace
\Bigr\rbrack m_{\tilde N_i}\Bigr\rbrace\cr
x_{\tilde a_m}^{\tilde g}=&{{m_{\tilde a_m}^2}\over
{m_{\tilde g}^2}}\cr
x_{\tilde a_m}^{\tilde N_i}=&{{m_{\tilde a_m}^2}\over
{m_{\tilde N_i}^2}}\cr
T_{m_ai}=&{{m_a}\over{m_Z\cos\beta}}N_{i3}\cr
T_i^{aL}=&e_a\sin2\Theta_WN'_{i1}-(1+2e_a\sin^2\Theta_W)N'_{i2}\cr
T_i^{aR}=&-\lbrace e_a\sin2\Theta_WN'_{i1}-2e_a\sin^2\Theta_W
N'_{i2}\rbrace\cr
}$$
where $a$\ runs over all 3 generations. Note that in the SM
the up quarks are running in the loop, whereas in the MSSM 
the scalar down quarks are.
Also note that $T_i^{sL,R}=T_i^{dL,R}$. $\cos\beta$\ can
be extracted from $\tan\beta$, $N_{ij}$\ and $N'_{ij}$\ 
are the diagonalizing angles of the neutralinos as
defined in eq.(A.20), eq.(A.23) and shown in Fig.24 in
[19]. I take them to be real and put all unknown SUSY
phases  into  
$K_{abL,R}^{\tilde N}$. The functions $A_a^{SM}$\ and
$B_a^{SM}$\ are taken from [23]. 
When the mixing of the scalar down quarks is neglected,
the functions $A_a^{\tilde g}$\ and
$B_a^{\tilde g}$\ agree with [9] (see Appendices A+B)
up to a relative minus
sign of the term proportional to the gluino mass.  
\hfill\break\indent
$\eta_{QCD}$\ is
a QCD factor obtaining the structure functions of the
Kaons, pions as well as their masses with
dimension $1/GeV$\ and given in 
eq.(14) in [9]. There it was taken $f_K=f_\pi$\ leading
to $\eta_{QCD}\simeq 10.8$\ and to a bit smaller value
of $\simeq 8.3$\ for $f_K=1.27 f_\pi$\ [27] for the masses
taken there. $f\simeq 1/6$\ and 
$m_Q$\ is the constituent $u,d$\ quark mass
taken to be $0.3$\ GeV, $m_s=0.5$\ GeV, $m_b=4.5$\ GeV,
$m_c=1.3$\ GeV and $m_t=180$\ GeV.
\hfill\break\indent
The functions $A_F,\ A_G,\ B_F,\ B_G,\ \tilde B_F$\ and $\tilde B_G$\
are given in the Appendix B. $C_{2G}=N$\ and $C_{2F}=
(N^2-1)/2N$\ are the Casimir operators of the adjoint
and fundamental $SU(N)$\ representation respectively,
obtained by the relations $T^bT^aT^b=\lbrack -{1\over 2}
C_{2G}+C_{2F}\rbrack T^a$\ and $f^{abc}T^bT^c={i\over 2}
C_{2G}T^a$.
\hfill\break\indent  
Before I discuss the results I have to make some comments.
The final results of the calculation of the SUSY penguin 
and self energy diagrams
are finite after summation. The infinite terms proportional
to $C_{2F}$\ and $C_{2G}$\ are cancelled seperately. Whereas
the $C_{2G}$\ infinite terms are cancelled by summation over
both penguin diagrams are 
the $C_{2F}$\ ones cancelled by summation over the penguin diagram
with two scalars and the self energy diagrams.
The calculation for the penguin diagrams with neutralinos
in the loop is similiar to the $C_{2F}$\ terms of the diagrams
with the gluino in the loop.
\hfill\break\indent 
In the literature it often can be found that the infinities
are cancelled by the GIM mechanisme (that is making use of
$\sum\limits_{a=1-3}K^\ast_{a1}K_{a2}=0$).
I find it not legitimate to do so since the model of the KM
matrix is a model within the SM, which is a renormalizable
model and therefore its divergencies should be removed by counter
terms or by summation of all possible diagrams.
\hfill\break\indent
It is also well known that SUSY has new CP violation phases
in the supersymmetric breaking sector (gaugino masses, scalar
masses, $A,\ \mu$\ term, vev's etc), which are strongly bounded
by the electric dipole moment of the neutron (EDMN)  
to be of the order of $10^{-2}-10^{-3}$\ [24, 25] or the 
SUSY masses are heavier than several TeV's [26]. 
In eq.(7) the SUSY phase $\Phi_S$\ only comes in when
$K^{\tilde g,\tilde N_i}_L$\ is multiplied
by $K^{\tilde g,\tilde N_i}_R$\
(the difference between the left and right
part lies in a relative minus sign of the SUSY phase, see eq.(4) in [9]),
that is the terms,
which are proportional to $\tilde K^a_{m1}
\tilde K^a_{m2}$. 
For example for flavour non diagonal
couplings of the gluino to the left and right handed down quarks
and their superpartners 
and after summation of the scalar down quark mass
eigenstates the gluino mass term of $B_a^{\tilde g}$\
in eq.(7) is proportional to
$\sin(2\Phi_S)
\sin2\Theta_a(G(x_{\tilde a_2}^{\tilde g})-
G(x_{\tilde a_1}^{\tilde g}))$\ and therefore only
contributes for non neglectable mixing of the scalar
down quarks and for a non neglectable SUSY phase $\Phi_S$. 
In this paper I assume that CP violation occurs in the
supersymmetric version of the KM matrix as well as
in a SUSY phase but take $\sin(2\Phi_S)\le 10^{-3}$,
and  the SUSY masses to be lower than the TeV range.
\hfill\break\indent
In the SM model we have
 $\xi_{SM}\approx -fc_2s_2s_3\sin\delta(A_2^{SM}-
A_3^{SM})/c_1c_3(A_1^{SM}-A_2^{SM})\simeq -4.34\times 10^{-4}$\
for a top quark mass of $180$\ GeV and the other quark masses
taken as given above. For the KM angles I take $s_1=0.22$, 
$s_2=0.095$, $s_3=0.05$\ and $\sin\delta=0.2$\ [28], which
reproduces approximately the results of [23]
(when the quark masses are taken as there and without the factor $f$). 
As was pointed out there
the contribution from $B_a^{SM}$\ can roughly be neglected even
for higher values of the top quark masses
and is less than $7$\% ($\xi_{SM}\simeq
-4.69\times 10^{-4}$\ when included, which I will in the final
results).
\hfill\break\indent
As mentioned above I assumed that both couplings
of the gluino and the neutralinos to the left and right 
handed down quarks and their
superpartners are flavour non diagonal
and to be of the same order, that is
I take $K_{abR}^{\tilde g}\approx
K_{abR}^{\tilde N}= e^{-i\Phi_S}K_{ab}$\
and $K_{abL}^{\tilde g}\approx K_{abL}^{\tilde N}=e^{+i\Phi_S}K_{ab}$,
where $K_{ab}$\ is the
supersymmetric version of the KM matrix
diagonalizing the scalar down quark matrix.
\hfill\break\indent  
The $\xi$\ parameter in SUSY alone has a factor of
$\varepsilon^2\sin\delta_{SKM}$\ compared to the SM one
of $c_2s_2s_3\sin\delta/c_1c_3$ leading to the same order
as in the SM if $\varepsilon\simeq 0.1$\ and
$\delta_{SKM}\simeq\delta$, where $\delta_{SKM}$\ is the 
supersymmetric version of the phase of the KM matrix and
$\varepsilon=\tilde s_a$\ with $\tilde s_a$\ the supersymmetric
version of the angles in the KM matrix (see e.g. [15,16,20]).  
An enhancement of $\varepsilon$\
can be compensated by a diminishing of $\delta_{SKM}$. I therefore
take the same values for the SUSY KM parameters as in the SM
and compare the results obtained from eq.(7) with the one
obtained via the SM without the gluino and neutralino
contributions. 
\hfill\break\indent
I include the mixing of all generations of the 
scalar down quarks since as I have shown in [29] 
the mixing angles might also become important in the second
generation. It can not be neglected here since the finite
result in eq.(7) was obtained by an expansion in the 
down quark masses as shown in Appendix A; 
furthermore for large values of $\tan\beta\gg 1$\ I obtain
$\cos^2\Theta_b\approx 1/\sqrt{2}$.
\hfill\break\indent
\vskip.2cm\noindent
{\bf III. DISCUSSIONS}\vskip.2cm
I now present those contributions  for different values of 
$\tan\beta$\ and the symmetry-breaking scale $m_S$.
As input parameter I take the quark masses as given above,
for $\sin^2\Theta_W=0.2323$, $\alpha = 1/137$\ and 
for the strong coupling
constant $\alpha_s=0.1134$. Furthermore for not having too many
parameters I use the well known GUT relations
$m_{g_1}={5\over 3}m_{g_2}\tan^2\Theta_W$\ and $m_{g_2}=
(g_2/g_s)^2m_{\tilde g}$\ between the $U(1)$, $SU(2)$\ and
$SU(3)$ gaugino and gluino masses.
\hfill\break\indent
In Fig. 2, I show the ratio of the neutralino contribution 
and the SM model contribution to
the direct CP violation parameter
 $\xi^{\rm \tilde N+SM}/\xi^{\rm SM}$\ and 
compare them with the gluino contribution for three
different values of $\tan\beta=1,\ 10$\ and $50$
\footnote{$^2$}{Such high values for $\tan\beta$\ are preferred
in models, which require the Yukawa couplings $h_t,\ h_b\ 
{\rm and}\ h_\tau$\ to meet at one point at the unification scale  
[31]}
and a gaugino mass of $m_{g_2}=200$\ GeV. 
Here I have
taken $\sin(2\Phi_S)=10^{-3}$. 
The ratio of $\xi^{\rm SM+\tilde N}/\xi^{\rm SM}$\
corresponds to the ratio $\epsilon'^{\rm SM+\tilde N}/
\epsilon'^{\rm SM}\approx (\epsilon'/\epsilon)^{\rm SM+\tilde N}
/(\epsilon'/\epsilon)^{\rm SM}$\ if we assume that
the SUSY contribution to $\epsilon_K$\ is neglectable. 
\hfill\break\indent
As we can see the neutralino
contribution is more important than the gluino one for small
values of $m_S\le 400$\ GeV. For higher values the neutralino
contribution is neglectable compared to the gluino and SM one. In the 
gluino case the most important term in eq.(7) is the one with
the gluino mass ($m_{\tilde g}=722$\ GeV with the relation 
given above) for $\sin(2\Phi_S)\not=0$. For $\sin(2\Phi_S)=0$\
the gluino contribution is totally neglectable compared to the
SM one, whereas the neutralino contribution is almost the same
as before. I therefore only present the
results with the upper limit for the SUSY CP violating phase. 
The gluino contribution becomes more important for higher
values of $\tan\beta$, whereas it is the opposite case for
the neutralino contribution. For small values of $m_S$\ the
SUSY contribution to $\xi$\ even becomes negative.
\hfill\break\indent
Smaller values as well as much larger values of the gaugino
mass $m_{g_2}$\ in general lead to a gluino contribution 
almost indistinguishable from the SM one, whereas the
neutralino contribution becomes somewhat bigger or smaller dependant 
on $\tan\beta$. The shapes of the figure however are not
changed.
Changing the sign of the $\mu$\ parameter (the bilinear Higgs
mass term in the superpotential), which enters the mass matrices
of the neutralino and scalar quark masses, affects the sign
of the $\xi$\ parameter in the neutralino and gluino case,
leading to results above the ratio $1$.
However here only for large $\tan\beta$\ and for the gluino
case the contribution is non neglectable to the SM one.
Smaller values of $c=-1$\ [15, 16] the parameter,
which enters in the one loop corrections to the scalar
down quark masses affects the results only slightly in the cases
presented here.
\hfill\break\indent
As a result I have that the neutralino contribution becomes
more important than the gluino one for small values of 
$m_S$. Unfortunately this is only for a small range of $m_S$\
since for $m_S$\ smaller than about $300$\ GeV 
one or both of the ligthest mass eigenvalues
of the scalar bottom and scalar top quark mass becomes negative
and so has to be discarded to avoid colour breaking.
\hfill\break\vskip.12cm\noindent
{\bf IV. CONCLUSIONS}\vskip.12cm
In this paper I presented the contribution of the neutralinos
and gluino to the direct CP violating parameter $\xi$. I have
shown that for small masses of the SUSY breaking scalar
mass $m_S\le 400$\ GeV the neutralino contribution becomes
more important than the gluino one. For a soft CP violation
phase of zero the gluino contribution is totally neglectable
compared to the SM contribution for a high gluino mass,
whereas the neutralinos one remains important for a scalar
mass smaller than $400$\ GeV. In the calculation I included the
mixing of the neutralinos and 
of the scalar partner of the left and right handed 
down quarks.
\hfill\break\indent
Although it was generally believed that the SUSY contribution
to the indirect CP violating parameter $\epsilon_K$\ can be
neglected the authors in [15] found non neglectable interference
between penguin and box diagrams for values of 
$x=m_{\tilde g}^2/m_S^2\approx 1$ of the gluino diagrams.
Since for a certain range of the parameter $m_S$\ the neutralino
penguin diagrams cannot be neglected as I have shown in this
paper I conclude that this might be also true for the box
diagrams [18]. This is further supported by the fact that
in the range mentioned above the neutralino contribution
to the mass difference in the $B_d^0$\ system cannot be
neglected compared to the gluino contribution as we have
shown in [16].
\hfill\break\indent
Finally I want to mention that the chargino and Higgs
contribution to $\epsilon'/\epsilon$\ can enhance the
SM by at most $40$\% as was shown in [14]. For scalar
masses $m_S$\ higher than $400$\ GeV their contribution
therefore becomes certainly the most important one, whereas
for smaller values all particles within the MSSM have to
be taken into account. 
\hfill\break\vskip.1cm\noindent
{\bf V. ACKNOWLEDGMENTS}\vskip.12cm 
\noindent
I like to thank C. Hamzaoui for useful discussions
and M.A. Doncheski for careful reading of the manuscript
and useful comments.
The figures were done with the 
program PLOTDATA from TRIUMF and I used the CERN-Library to 
diagonalize the neutralino mass matrix.
\hfill\break\indent
This work was partially funded by funds from the N.S.E.R.C. of
Canada and les Fonds F.C.A.R. du Qu\'ebec. 
\hfill\break\vskip.12cm\noindent
{\bf VI. APPENDIX A}\vskip.12cm
In this Appendix I present the complete calculations of
the Feynman diagrams as shown in Fig.1 with 
flavour non diagonal couplings of the gluino to the
left and right handed quarks and their scalar partners.
\hfill\break\indent
The first Feynman diagram of Fig.1 leads to the following
expression after dimensional integration:
$$\eqalignno{iM_1=&+{{2g_s^3}\over{(4\pi)^2}}
\lbrack -{1\over 2}C_{2G}+C_{2F}\rbrack\int\limits_0^1
d\alpha_1\int\limits_0^{1-\alpha_1}d\alpha_2
\Bigl\lbrace
\bigl\lbrack {1\over\epsilon}-\gamma+\log(4\pi\mu^2)-
\log(F_{\tilde g\tilde a_m})\bigr\rbrack
\cr&
\overline u_dT^a\gamma_\mu
(K^{\ast\tilde g}_{a1L}
K^{\tilde g}_{a2L}\tilde K_{m1}^{a2}P_L+
K^{\ast\tilde g}_{a1R}K^{\tilde g}_{a2R}\tilde K_{m2}^{a2}P_R)
u_s
\cr&
+\overline u_dT^a\bigl\lbrack\rlap/{\tilde p}
(K^{\ast\tilde g}_{a1L}K^{\tilde g}_{a2L}\tilde K_{m1}^{a2}P_L+
K^{\ast\tilde g}_{a1R}K^{\tilde g}_{a2R}\tilde K_{m2}^{a2}P_R)
\cr&
-m_{\tilde g}(K^{\ast\tilde g}_{a1R}K^{\tilde g}_{a2L}P_L
+K^{\ast\tilde g}_{a1L}K^{\tilde g}_{a2R}P_R)\tilde K^a_{m1}
\tilde K^a_{m2}\bigr\rbrack u_s
(p_1+p_2-2\tilde p)_\mu/F_{\tilde g\tilde a_m}
\Bigr\rbrace\epsilon^{\ast\mu}_g\cr
\tilde p=&p_1\alpha_1+p_2\alpha_2
\cr
F_{\tilde g\tilde a_m}=&m_{\tilde g}^2-(m_{\tilde g}^2-
m_{\tilde a_m}^2)(\alpha_1+\alpha_2)-p_1^2\alpha_1(1-
\alpha_1-\alpha_2)
-p_2^2\alpha_2
(1-\alpha_1-\alpha_2)
\cr&
-q^2\alpha_1\alpha_2&(A.1)
\cr}$$
$\epsilon=2-d/2$\ and $q=p_1-p_2$.
\hfill\break\indent
For the second diagram in Fig.1 the result is:
$$\eqalignno{iM_2=&+{{2g_s^3}\over{(4\pi)^2}} 
{1\over 2}C_{2G}\int\limits_0^1
d\alpha_1\int\limits_0^{1-\alpha_1}d\alpha_2
\Bigl\lbrace\bigl\lbrack {1\over\epsilon}-1-\gamma+
\log(4\pi\mu^2)-\log(F_{\tilde a_m\tilde g})+
{{m_{\tilde g}^2}\over{F_{\tilde a_m\tilde g}}}\bigr\rbrack
\cr&
\overline u_dT^a\gamma_\mu
(K^{\ast\tilde g}_{a1L}
K^{\tilde g}_{a2L}\tilde K_{m1}^{a2}P_L+
K^{\ast\tilde g}_{a1R}K^{\tilde g}_{a2R}\tilde K_{m2}^{a2}P_R)
u_s
\cr&
+\overline u_dT^a\bigl\lbrace(\rlap/p_2-\rlap/{\tilde p})\gamma_\mu
(\rlap/p_1-\rlap/{\tilde p})(K^{\ast\tilde g}_{a1L}
K^{\tilde g}_{a2L}\tilde K_{m1}^{a2}P_L+
K^{\ast\tilde g}_{a1R}K^{\tilde g}_{a2R}\tilde K_{m2}^{a2}P_R)
\cr&
-m_{\tilde g}\lbrack(\rlap/p_2-\rlap/{\tilde p})\gamma_\mu+
\gamma_\mu(\rlap/p_1-\rlap/{\tilde p})\rbrack
(K^{\ast\tilde g}_{a1R}K^{\tilde g}_{a2L}P_L
+K^{\ast\tilde g}_{a1L}K^{\tilde g}_{a2R}P_R)\tilde K^a_{m1}
\tilde K^a_{m2}\bigr\rbrace 
\cr&
u_s/F_{\tilde a_m\tilde g}
\Bigr\rbrace
\epsilon^{\ast\mu}_g&(A.2)
\cr}$$
Finally for the self energy diagrams the results are:
$$\eqalignno{iM_3=&-{{2g_s^3}\over{(4\pi)^2}}C_{2F}
\int\limits_0^1\Bigl\lbrace\lbrack {1\over\epsilon}-
\gamma+\log(4\pi\mu^2)-\log(H_{\tilde g\tilde a_m}^{p_2}
\rbrack
\cr&
\overline u_dT^a\bigl\lbrace\rlap/p_2\alpha_1
(K^{\ast\tilde g}_{a1L}K^{\tilde g}_{a2L}\tilde K_{m1}^{a2}P_L
+K^{\ast\tilde g}_{a1R}K^{\tilde g}_{a2R}\tilde K_{m2}^{a2}P_R)
\cr&
-m_{\tilde g}(K^{\ast\tilde g}_{a1R}K^{\tilde g}_{a2L}P_L
+K^{\ast\tilde g}_{a1L}K^{\tilde g}_{a2R}P_R)\tilde K^a_{m1}
\tilde K^a_{m2}\bigr\rbrace
{{\rlap/p_2+m_s}\over{p_2^2-m_s^2}}\gamma_\mu u_s
\Bigr\rbrace\epsilon^{\ast\mu}_g&(A.3)
\cr
iM_4=&-{{2g_s^3}\over{(4\pi)^2}}C_{2F}
\int\limits_0^1\Bigl\lbrace\lbrack {1\over\epsilon}-
\gamma+\log(4\pi\mu^2)-\log(H_{\tilde g\tilde a_m}^{p_1}
\rbrack
\cr&
\overline u_dT^a\gamma_\mu{{\rlap/p_1+m_d}\over{p_1^2-m_d^2}}
\bigl\lbrace\rlap/p_1\alpha_1
(K^{\ast\tilde g}_{a1L}K^{\tilde g}_{a2L}\tilde K_{m1}^{a2}P_L
+K^{\ast\tilde g}_{a1R}K^{\tilde g}_{a2R}\tilde K_{m2}^{a2}P_R)
\cr&
-m_{\tilde g}(K^{\ast\tilde g}_{a1R}K^{\tilde g}_{a2L}P_L
+K^{\ast\tilde g}_{a1L}K^{\tilde g}_{a2R}P_R)\tilde K^a_{m1}
\tilde K^a_{m2}\bigr\rbrace u_s
\Bigr\rbrace\epsilon^{\ast\mu}_g&(A.4)
\cr
H_{\tilde g\tilde a_m}^{p_q}=&m_{\tilde g}^2-(m_{\tilde g}^2-
m_{\tilde a_m})\alpha_1-p_q^2\alpha_1(1-\alpha_1)
\cr}$$
Neglecting all quark masses 
(that is with $p_{1,2}^2=0=q^2$) the $\gamma_\mu$
term is identical to zero
after summation over all four terms. For the $C_{2G}$\ the
result is:
$$\int\limits_0^1d\alpha_1\int\limits_0^{1-\alpha_1}d\alpha_2
\bigl\lbrace{1\over\epsilon}-1-\gamma
+\log(4\pi\mu^2)-\log(F_{\tilde a_m\tilde g})
+{{m_{\tilde g}^2}\over{F_{\tilde a_m\tilde g}}}-{1\over\epsilon}
+\gamma-\log(4\pi\mu^2)+\log(F_{\tilde g\tilde a_m})
\bigr\rbrace\equiv 0
\eqno(A.5)$$
And for the $C_{2F}$\ term I obtain:
$$\eqalignno{&\int\limits_0^1d\alpha_1
\int\limits_0^{1-\alpha_1}d\alpha_2
\overline u_dT^a
(K^{\ast\tilde g}_{a1L}K^{\tilde g}_{a2L}\tilde K_{m1}^{a2}P_L
+K^{\ast\tilde g}_{a1R}K^{\tilde g}_{a2R}\tilde K_{m2}^{a2}P_R)
u_s
\cr&
\bigl\lbrace {1\over\epsilon}-\gamma+
\log(4\pi\mu^2)-\log(F_{\tilde g\tilde a_m})\bigr\rbrace
\cr&
-\int\limits_0^1d\alpha_1\overline u_dT^a\bigl\lbrace\rlap/p_2
(K^{\ast\tilde g}_{a1L}K^{\tilde g}_{a2L}\tilde K_{m1}^{a2}P_L
+K^{\ast\tilde g}_{a1R}K^{\tilde g}_{a2R}\tilde K_{m2}^{a2}P_R)
{{\rlap/p_2+m_s}\over{p_2^2-m_s^2}}\gamma_\mu
\cr&
+\gamma_\mu{{\rlap/p_1+m_d}\over{p_1^2-m_d^2}}\rlap/p_1
(K^{\ast\tilde g}_{a1L}K^{\tilde g}_{a2L}\tilde K_{m1}^{a2}P_L
+K^{\ast\tilde g}_{a1R}K^{\tilde g}_{a2R}\tilde K_{m2}^{a2}P_R)
\bigr\rbrace u_s
\cr&
\times\alpha_1\bigl\lbrace {1\over\epsilon}-\gamma
+\log(4\pi\mu^2)-\log(H_{\tilde g\tilde a_m})\bigr\rbrace
\equiv 0
&(A.6)
\cr}$$
The term proportional to the gluino mass in eq.(A.3) and eq.(A.4)
cancels after summation. To obtain eq.(A.6) and the cancellation
of the gluino mass term of the self energy diagrams for zero quark
masses we have to use the relations $\overline u_dP_{L,R}
\rlap/p_2=m_d\overline u_dP_{R,L}$, $\rlap/p_2P_{L,R}u_s=
m_sP_{R,L}u_s$\ and $\rlap/p_{1,2}\rlap/p_{1,2}=m^2_{s,d}$.
Since the summation of eq.(A.1-4) gives a zero result we have
to expand the functions $F_{\tilde g\tilde a_m}$, $F_{\tilde a_m
\tilde g}$\ and $H_{\tilde g\tilde a_m}^{p_q}$\ with respect
to $p^2_1$, $p_2^2$\ and $q^2$. For the further calculation
I strongly make use of similar relations as presented in eq.(A.6) and
eq.(A.7) in [30]. Furthermore we have:
$$\eqalignno{
\overline u_di\sigma_{\mu\nu}q^\nu(m_dP_L+m_sP_R)u_s=&
\overline u_d\lbrace\rlap/p_1p_{1\mu}+\rlap/p_2p_{2\mu}
+\rlap/p_1p_{2\mu}+\rlap/p_2p_{1\mu}-2\rlap/p_2\gamma_\mu
\rlap/p_1
\cr&
-(p_1^2+p_2^2)\rbrace P_Lu_s&(A.7)\cr}$$
$$\rlap/p_1\gamma_\mu\rlap/p_2=2(\rlap/p_1p_{2\mu}+
\rlap/p_2p_{1\mu})+(q^2-p_1^2-p_2^2)\gamma_\mu-
\rlap/p_2\gamma_\mu\rlap/p_1\eqno(A.8)$$
As a final result I obtain:
$$\eqalignno{iM=&+{{2g_s^3}\over{(4\pi)^2}}{1\over
{m_{\tilde g}^2}}\overline u_dT^a\lbrack {1\over 2}
C_{2G}M_G+C_{2F}M_F\rbrack u_s&(A.9)\cr
M_G=&\int\limits_0^1d\alpha_1\Bigl\lbrace
{1\over 6}(q^2\gamma_\mu-\rlap/qq_\mu)
\lbrack 2-3\alpha_1^2-3\alpha_1(1-\alpha_1)\rbrack
\cr&
(K^{\ast\tilde g}_{a1L}K^{\tilde g}_{a2L}\tilde K_{m1}^{a2}P_L
+K^{\ast\tilde g}_{a1R}K^{\tilde g}_{a2R}\tilde K_{m2}^{a2}P_R)
\cr&
-i\sigma_{\mu\nu}q^\nu{1\over 2}\alpha_1(1-\alpha_1)
\bigl\lbrack m_d
(K^{\ast\tilde g}_{a1L}K^{\tilde g}_{a2L}\tilde K_{m1}^{a2}P_L
+K^{\ast\tilde g}_{a1R}K^{\tilde g}_{a2R}\tilde K_{m2}^{a2}P_R)
\cr&
+m_s(K^{\ast\tilde g}_{a1L}K^{\tilde g}_{a2L}\tilde K_{m1}^{a2}P_R
+K^{\ast\tilde g}_{a1R}K^{\tilde g}_{a2R}\tilde K_{m2}^{a2}P_L)
\bigr\rbrack
\cr&
+i\sigma_{\mu\nu}q^\nu(1-\alpha_1)m_{\tilde g}
(K^{\ast\tilde g}_{a1R}K^{\tilde g}_{a2L}P_L
+K^{\ast\tilde g}_{a1L}K^{\tilde g}_{a2R}P_R)\tilde K^a_{m1}
\tilde K^a_{m2}\Bigr\rbrace/D_{\tilde g\tilde a_m}
\cr M_F=&\int\limits_0^1d\alpha_1\Bigl\lbrace
{1\over 6}(q^2\gamma_\mu-\rlap/qq_\mu)
(K^{\ast\tilde g}_{a1L}K^{\tilde g}_{a2L}\tilde K_{m1}^{a2}P_L
+K^{\ast\tilde g}_{a1R}K^{\tilde g}_{a2R}\tilde K_{m2}^{a2}P_R)
\alpha_1^3
\cr&
+i\sigma_{\mu\nu}q^\nu{1\over 2}\alpha_1^2(1-\alpha_1)
\bigl\lbrack m_d
(K^{\ast\tilde g}_{a1L}K^{\tilde g}_{a2L}\tilde K_{m1}^{a2}P_L
+K^{\ast\tilde g}_{a1R}K^{\tilde g}_{a2R}\tilde K_{m2}^{a2}P_R)
\cr&
+m_s(K^{\ast\tilde g}_{a1L}K^{\tilde g}_{a2L}\tilde K_{m1}^{a2}P_R
+K^{\ast\tilde g}_{a1R}K^{\tilde g}_{a2R}\tilde K_{m2}^{a2}P_L)
\bigr\rbrack
\cr&
-i\sigma_{\mu\nu}q^\nu\alpha_1(1-\alpha_1)m_{\tilde g}
(K^{\ast\tilde g}_{a1R}K^{\tilde g}_{a2L}P_L
+K^{\ast\tilde g}_{a1L}K^{\tilde g}_{a2R}P_R)\tilde K^a_{m1}
\tilde K^a_{m2}\Bigr\rbrace/D_{\tilde g\tilde a_m}
\cr
D_{\tilde g\tilde a_m}=&1-(1-x_{\tilde a_m}^{\tilde g})\alpha_1\cr
x_{\tilde a_m}^{\tilde g}=&{{m_{\tilde a_m}^2}\over{m_{\tilde g}^2}}
\cr}$$
After Feynman integration eq.(A.7) leads to the results
presented in eq.(7).
The final functions after Feynman integration are shown in
Appendix B. Up to a relative minus sign
\footnote{$^3$}{The relative minus sign is
the relative sign between the couplings of
the gluino to the left handed down quarks
and their superpartners as explained in eq.(C89)
in [8], which was not taken into account in [9]} of the term proportional
to the gluino mass the results presented in [9] are reproduced
by the following replacements: $A_G=A/3$, $A_F=B/6$, $B_G=C$,
$B_F=D/2$, $\tilde B_G=E$\ and $\tilde B_F=2C$, where
$A_G,\ A_F,\ B_G,\ B_F,\ \tilde B_G,\ \tilde B_F$\
are given in eq.(B.1-6) and $A,\ B,\ C,\ D$\ and $E$\
are the functions given in eq.11 of [9]. I therefore cannot
confirm the statement of [17], that the authors in [9]
neglected a crucial term. The function $F(x_j)$\ of eq.7
in [17] is reproduced after Feynman integration of the
term $2-3\alpha_1^2$\ in $A_G$\ however they omitted
the term $-3\alpha_1(1-\alpha_1)$\ there, which cancels
the $\alpha_1^2$\ leaving the term $2-3\alpha_1$\ and thus
reproducing the function $A$\ of [9]. 
\hfill\break\vskip.12cm\noindent
{\bf VI. APPENDIX B}\vskip.12cm
$$\eqalignno{
A_G(x_{\tilde a_m}^{\tilde g})=&{1\over 6}\int\limits_0^1
d\alpha_1\bigl\lbrack 2-3\alpha_1^2-3\alpha_1(1-\alpha_1)
\bigr\rbrack/ D_{\tilde g \tilde a_m}\cr
=&{1\over{(1-x_{\tilde a_m}^{\tilde g})^2}}{1\over 2}
\bigl\lbrace 1-x_{\tilde a_m}^{\tilde g}+{1\over 3}(1+2
x_{\tilde a_m}^{\tilde g})
\log(x_{\tilde a_m}^{\tilde g})\bigr\rbrace&(B.1)
\cr
A_F(x_{\tilde a_m}^{\tilde g})=&{1\over 6}\int\limits_0^1
d\alpha_1\alpha_1^3/D_{\tilde g \tilde a_m}\cr
=&{1\over{(1-x_{\tilde a_m}^{\tilde g})^4}}{1\over 36}
\bigl\lbrace -11+18 x_{\tilde a_m}^{\tilde g}
-9x_{\tilde a_m}^{\tilde g\ 2}
+2x_{\tilde a_m}^{\tilde g\ 3}-
6\log(x_{\tilde a_m}^{\tilde g})\bigr\rbrace&(B.2)
\cr
B_G(x_{\tilde a_m}^{\tilde g})=&{1\over 2}\int\limits_0^1
d\alpha_1\alpha_1(1-\alpha_1)/D_{\tilde g \tilde a_m}
\cr
=&{1\over{(1-x_{\tilde a_m}^{\tilde g})^3}}{1\over 4}
\bigl\lbrace 1-x_{\tilde a_m}^{\tilde g\ 2}+
2x_{\tilde a_m}^{\tilde g}\log(
x_{\tilde a_m}^{\tilde g})\bigr\rbrace&(B.3)\cr
B_F(x_{\tilde a_m}^{\tilde g})=&{1\over 2}\int\limits_0^1
d\alpha_1\alpha_1^2(1-\alpha_1)/D_{\tilde g \tilde a_m}\cr
=&{1\over{(1-x_{\tilde a_m}^{\tilde g})^4}}{1\over 12}\bigl\lbrace
2+3x_{\tilde a_m}^{\tilde g}-
6x_{\tilde a_m}^{\tilde g\ 2}+x_{\tilde a_m}^{\tilde g\ 3}
+6x_{\tilde a_m}^{\tilde g}\log(x_{\tilde a_m}^{\tilde g})\bigr\rbrace
&(B.4)\cr
\tilde B_G(x_{\tilde a_m}^{\tilde g})=&\int\limits_0^1d\alpha_1
(1-\alpha_1)/D_{\tilde g \tilde a_m}
={1\over{(1-x_{\tilde a_m}^{\tilde g})^2}}
\bigl\lbrace 1-x_{\tilde a_m}^{\tilde g}
+x_{\tilde a_m}^{\tilde g}\log(x_{\tilde a_m}^{\tilde g})\bigr\rbrace
&(B.5)\cr
\tilde B_F(x_{\tilde a_m}^{\tilde g})=
&2 B_G(x_{\tilde a_m}^{\tilde g})&(B.6)\cr
\cr
}$$
\hfill\break\vskip.12cm\noindent
{\bf REFERENCES}\vskip.12cm
\item{[\ 1]}G.D. Barr et al (NA31 Collaboration), Phys.Lett.
{\bf 317B}(1993)233.
\item{[\ 2]}L.K. Gibbons et al (E731 Collaboration),
Phys.Rev.Lett.{\bf 70}(1993)1203.
\item{[\ 3]}F. Abe et al, Phys.Rev.Lett.{\bf 74}(1995)2626;
S. Abachi et al, ibid, 2632.
\item{[\ 4]}E.A. Pachos and Y.L. Wu, Mod.Phys.Lett.{\bf A6}
(1991)93.
\item{[\ 5]}J. Heinrich et al, Phys.Lett.{\bf 279B}(1992)140.
\item{[\ 6]}M. Ciuchini et al, Phys.Lett.{\bf 301B}(1993)263.
\item{[\ 7]}A.J. Buras, M. Jamin and M.E. Lautenbacher,
Nucl.Phys.{\bf B408}(1993)209.
\item{[\ 8]}H.E. Haber and G.L. Kane, Phys.Rep.{\bf 117}
(1985)75.
\item{[\ 9]}J.M. G\'erard, W. Grimus and A.Raychaudhuri,
Phys.Lett.{\bf 145B}(1984)400.
\item{[10]}J.M. G\'erard et al, Nucl.Phys.{\bf B253}(1985)93.
\item{[11]}M.Dugan, B. Grinstein and L.J. Hall, Nucl.Phys.
{\bf B255}(1985)413.
\item{[12]}A. Dannenberg, L.J. Hall and L. Randall, Nucl.Phys.
{\bf B271}(1986)574.
\item{[13]}E. Gabrielli and G.F. Giudice, Nucl.Phys.{\bf B433}
(1995)3.
\item{[14]}E. Gabrielli, A. Masiero and L. Silvestrini,
``Flavour changing neutral currents and CP violating
processes in generalized supersymmetric theories'',
hep-ph/9509379.
\item{[15]}G. Couture and H. K\"onig,
Z.Phys.{\bf C69}(1996)499.
\item{[16]}G. Couture and H. K\"onig, ``Neutralino
contribution to the mass difference of the 
$B_d^0-\overline B_d^0$'', hep-ph/9511234 to be published
in Z.Phys.{\bf C}.
\item{[17]}S. Bertolini, F. Borzumati and A. Masiero, 
Nucl.Phys.{\bf B294}(1987)321.
\item{[18]}H. K\"onig, work in progress.
\item{[19]}J.F. Gunion and H. Haber, Nucl.Phys.{\bf
B272}(1986); Erratum-ibid.{\bf B402}(1993)567.
\item{[20]}M.J. Duncan, Nucl.Phys.{\bf B221}(1983)285.
\item{[21]}J.F. Donoghue, H.P. Nilles and D. Wyler, Phys.Lett.
{\bf 128B}(1983)55.
\item{[22]}G. Couture, C. Hamzaoui and H. K\"onig, Phys.Rev.
{\bf D52}(1995)1713.
\item{[23]}S.P. Chia, Phys.Lett.{\bf 130B}(1983)315.
\item{[24]}J. Ellis, S. Ferrara and D.V. Nanopoulos,
Phys.Lett.{\bf 114B}(1982)231;\hfill\break
 W. Buchm\"uller and D. Wyler,
Phys.Lett.{\bf 121B}(1983)321;\hfill\break
 J. Polchinski and M.B. Wise,
Phys.Lett.{\bf 125B}(1983)393;\hfill\break 
M. Dugan, B. Grinstein and L.J. Hall,
Nucl.Phys.{\bf B255}(1985)413.
\item{[25]} S. Bertolini and F. Vissani, Phys.Lett.{\bf 324B}(1994)
164;\hfill\break
T. Inui et al, Nucl.Phys.{\bf B449}(1995)49;\hfill\break
T. Kobayashi et al, Prog.Theor.Phys.{\bf 94}(1995)417;\hfill\break
S.A. Abel, W.N. Cottingham, I.B. Whittingham, "The electric
dipole moment of the neutron in the constrained minimally
supersymmetric standard model", hep-ph/9511326.
\item{[26]}G. Ecker, W. Grimus and H. Neufeld, Nucl.Phys.{\bf
B229}(1983)421;\hfill\break
 Y. Kizukiri and N. Oshimo, Phys.Rev.{\bf D45}
(1992)1806;ibid.{\bf D46}(1992)3025.
\item{[27]}K.R. Schubert, ``Flavour Oscillation'', IHEP-HD/87-3,
Maria Laach, Sep 1986.
\item{[28]}V. Barger et al, Phys.Lett.{\bf 194B}(1987)312.
\item{[29]}H. K\"onig, Z.Phys.{\bf C69}(1996)493. 
\item{[30]}H. K\"onig, Mod.Phys.Lett.{\bf A7}(1992)279.
\item{[31]}N. Polonsky, PhD thesis, hep-ph/9411378 and references
therein.
\hfill\break\vskip.12cm\noindent
{\bf FIGURE CAPTIONS}\vskip.12cm
\item{Fig.1}The penguin diagrams with scalar down quarks, 
gluino and neutralinos  within the loop.
\item{Fig.2} The ratios $\xi^{\rm SM+\tilde N}/\xi^{\rm SM}$\
and $\xi^{\rm SM+\tilde g}/\xi^{\rm SM}$\ for different
values of $\tan\beta=1$\ (solid line), $\tan\beta=10$\ (dashed
line) and $\tan\beta=50$\ (dot-dashed line) as a function
of $m_S$. $m_{g_2}=200$\ GeV ($m_{\tilde g}=722$\ GeV)
and $\mu=300$\ GeV. The soft
SUSY CP violating phase I took at its upper limit of
$\sin(2\Phi_S)=10^{-3}$. The lower curves for $m_S=300$\ GeV 
are for the neutralino contribution (except for $\tan\beta=50$, here the
neutralino one is the one closer to $1$).
\bye